\documentclass[12pt]{elsarticle}

\usepackage[]{amsmath}
\usepackage{graphicx}
\usepackage{epstopdf}
\usepackage{amssymb}
\usepackage[usenames]{color}
\definecolor{Red}{rgb}{1,0,0}

\usepackage{cases}
\usepackage[]{natbib}
\usepackage{verbatim}
\usepackage{lineno}
\usepackage{bm}
\usepackage{upgreek}

\usepackage{hyperref}
\usepackage{doi}
\usepackage{mathrsfs}
\usepackage{amsbsy}
\usepackage{xfrac}

\usepackage[a4paper,textwidth=16.5cm,textheight=23cm]{geometry}

\newcommand{\dd}{\mbox{\,d}}

\newcommand{\bn}{{\ensuremath{\bm{n}}}}
\newcommand{\bt}{{\ensuremath{\bm{t}}}}
\newcommand{\bN}{{\ensuremath{\bm{\nu}}}}
\newcommand{\bF}{{\ensuremath{\bm{f}}}}
\newcommand{\bT}{{\ensuremath{\bm{T}}}}
\newcommand{\bR}{{\ensuremath{\bm{\rho}}}}
\newcommand{\bI}{{\ensuremath{\pmb{\mathscr{I}}}}}
\newcommand{\bD}{{\ensuremath{\bm{\mathcal{D}}}}}
\newcommand{\cD}{{\ensuremath{{\mathcal{D}}}}}
\newcommand{\bE}{{\ensuremath{\bm{\mathcal{E}}}}}
\newcommand{\bIvol}{{\ensuremath{\pmb{\mathscr{I}}^{\mathrm{vol}}}}}
\newcommand{\pN}{{\ensuremath{\pmb{\mathscr{N}}}}}
\newcommand{\pT}{{\ensuremath{\pmb{\mathscr{T}}}}}
\newcommand{\Ndim}{{\ensuremath{N_{\mathrm{dim}}}}}

\newcommand{\lmin}{\ensuremath{l_{\min}}}

\journal{International Journal of Solids and Structures}

\begin{document}


\begin{frontmatter}
\title{ Elastic properties of isotropic discrete systems: connections between geometric structure and Poisson's ratio}

\author[ism]{Jan Eli\'{a}\v{s}\corref{cor1}}
\cortext[cor1]{Corresponding author}
\ead{jan.eliasj@vut.cz}

\address[ism]{Institute of Structural Mechanics, Faculty of Civil Engineering, Brno University of Technology, Veve\v{r}\'{i} 331/95, Brno, 60200, Czech Republic}

\begin{abstract}
The use of discrete material representation in numerical models is advantageous due to the straightforward way it takes into account material heterogeneity and randomness, and the discrete and orientated nature of cracks. Unfortunately, it also restricts the macroscopic Poisson's ratio and therefore narrows its applicability. The paper studies the Poisson's ratio of a~discrete model analytically. It derives theoretical limits for cases where the geometry of the model is completely arbitrary, but isotropic in the statistical sense. It is shown that the widest limits are obtained for models where normal directions of contacts between discrete units are parallel with the vectors connecting  these units. Any deviation from parallelism causes the limits to shrink.
A~comparison of the derived equations to the results of the actual numerical model is presented. It shows relatively large deviations from the theory because the fundamental assumptions behind the theoretical derivations are largely violated in systems with complex geometry. The real shrinking of the Poisson's ratio limit is less severe compared to that which is theoretically derived.
\end{abstract}

\begin{keyword}
lattice model, geometry, elasticity, Poisson's ratio, mesoscale, macroscopic characteristics
\end{keyword}

\end{frontmatter}

\section{Introduction}

Discrete modeling is a~well established technique in mechanics. It allows the explanation or prediction of the complex behavior of heterogeneous, cohesive or granular materials. The main advantages it offers are the straightforward representation of material random heterogeneity, the simple formulation of constitutive equations in vectorial form and also the direct consideration of discrete and oriented cracks. In contrast, the elastic behavior of these models still poses open challenges. Besides the minor issue of the inevitable boundary layer with different elastic properties \citep{Eli17}, the most serious problem lies in the inability of such models to exhibit Poisson's ratios greater than 1/3 for plane stress simplification and 1/4 for plane strain and three dimensional models (\citep{BatRot88,LiaCha-97,Eli17} or see Eq.~\ref{eq:lim_gamma_0}).

Recently, four remedies providing the full range of Poisson's ratio in discrete systems were presented. 
The first one \citep{AsaIto-15,AsaAoy-17} introduces artificial auxiliary stresses within an~iterative loop to achieve an~elastically homogeneous system with an~arbitrary Poisson's ratio. 
The other three methods are similar as all of them are based on estimation of tensorial stresses or strains. The tensorial quantity is always computed nonlocally, in some neighborhood of the contact or body. These methods take into account the lateral stresses (confinement effect). The second remedy \citep{CusRez-17} proposes constitutive model as a~function of the volumetric and deviatoric strain split. The third remedy \cite{CelLat-17} adds into the standard vectorial constitutive model terms accounting for the lateral stress. The fourth remedy \cite{RojZub-18} modifies distance between particles by integration tensorial strain over the body and evaluating body deformation. The stress oscillations caused by the heterogeneity of the material are unfortunately full or partially smeared out. Therefore, such models do not seem to be convenient for studying elastic behavior of highly heterogeneous structures at the mesoscale.

This paper is motivated by the author's long belief that the Poisson's ratio of discrete systems can be increased by changing the model geometry. Most of the papers published in this field use contact faces between discrete model units perpendicular to contact vectors \cite{GraAnt19,BolSai98}. Examples of models with skewed normals are mostly from the field of granular materials when non-spherical particle shapes are used \citep{RotBat91,GarLat-09,KilDon-19} but also static homogeneous models can be found \citep{YaoJia-16}. The assumption of perpendicularity is abandoned here allowing model of completely arbitrary geometry to be constructed. Poisson's ratio is then analyzed using strong assumptions about rotations and translations (Voigt's hypothesis) in the model adopted according to~\citep{KuhDadd-00}. It is proven here that abandoning  the perpendicularity leads only to shrinking of the interval of achievable Poisson's ratios.

The studied discrete system fills space continuously (without gaps or overlapping) with rigid bodies that possess translational ($\bm{u}$) and rotational ($\bm{\varphi}$) degrees of freedom. It is assumed that the system is isotropic -- arbitrary rotation of the domain does not change its geometrical properties in the statistical sense. The rigid bodies interact via contacts defined at their boundaries. The normal and tangential displacement discontinuities $\bm{\Delta}$ at these boundaries are dictated by rigid body kinematics and give rise to normal and tangential forces linearly dependent on the corresponding component of $\bm{\Delta}$. A~critical parameter governing the macroscopic Poisson's ratio is the ratio between tangential and normal contact stiffness, hereinafter denoted by $\alpha$. The parameter $\alpha$ must be non-negative, otherwise the system would exhibit negative stiffness and instability. One can find several examples of these models in literature~\citep{RezZho-17,FasOsk19,KanKim-14,AmaQia-18}.

The analytical derivation utilizes the equivalence of virtual work arising in the discrete system and continuum when they are subjected to equal straining. The Boltzmann continuum is used, and therefore the stress tensor must be symmetric (Boltzmann axiom). However, the discrete system yields a~non-symmetric stress tensor, as it is a~discrete instance of polar (Cosserat) continua instead \cite{RezCus16}. The virtual work equivalence is therefore accomplished with the help of the symmetrization of the tensor of elastic constants from the discrete model. 

\section{Normal and contact vector, volume}

\begin{figure}
\centering\includegraphics[width=12cm]{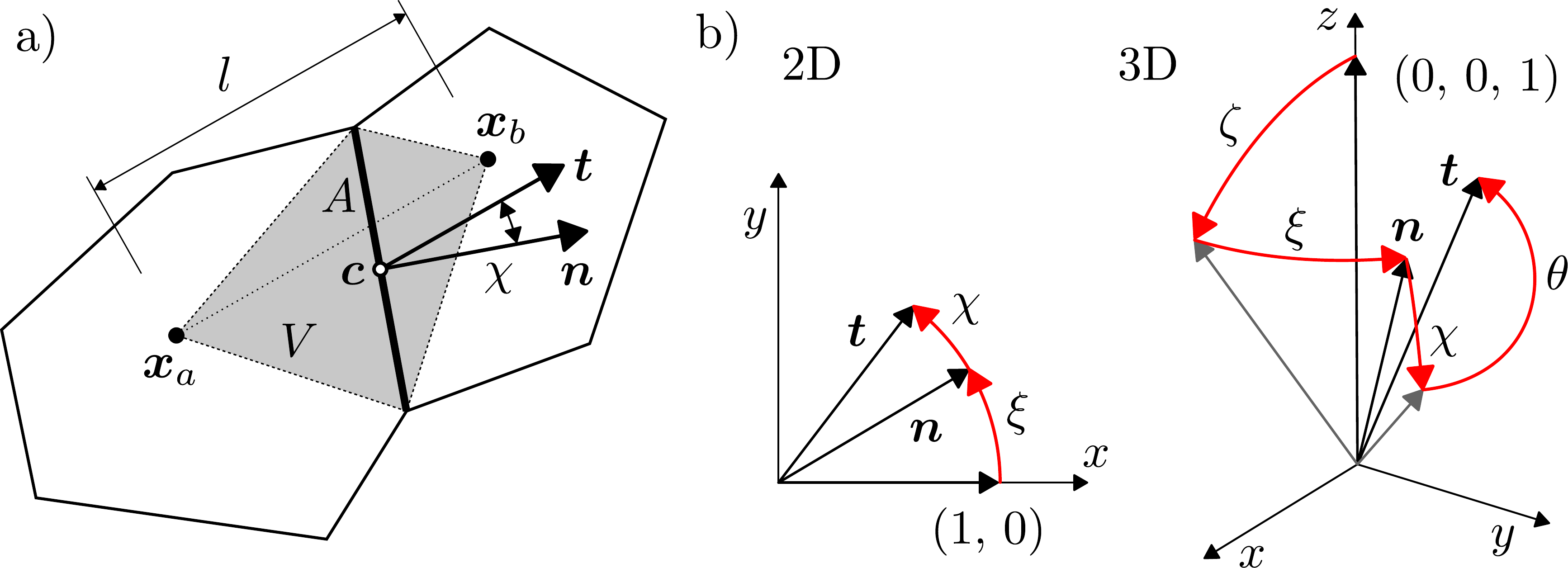}
\caption{a) Two dimensional rigid bodies in contact. The shaded area represents a~single mechanical element with normal vector $\bn$, contact vector $\bt$, area $A$, centroid $\bm{c}$, length $l$ and volume $V$. b) Normal and contact vector in two and three dimensions.} 
\label{fig:angles}
\end{figure}

The domain is divided into rigid bodies, each of which has degrees of freedom associated with the translations and rotations of its governing node, $\bm{x}_a$. The contact between two nodes $\bm{x}_a$ and $\bm{x}_b$ is provided by a~mechanical element with contact area $A$, length $l=\| \bm{x}_b-\bm{x}_a \|$, unit normal vector $\bn$ and contact vector $\bt=(\bm{x}_b-\bm{x}_a)/l$. The situation is depicted in Fig.~\ref{fig:angles}a in two dimensions.

We assume that the system has no directional bias, and that therefore all normal directions share the same probability of occurrence. The vector $\bm{n}$ is here defined in the Cartesian coordinate system by two angles, $\xi$ and $\zeta$
\begin{align}
\bn = \begin{cases} \left(\begin{array}{cc}\cos\xi & \sin\xi \end{array}\right) & \mathrm{in\ 2D} \\
\left(\begin{array}{ccc}\cos\xi \sin\zeta & \sin\xi \sin\zeta & \cos\zeta\end{array}\right) & \mathrm{in\ 3D}\end{cases}
\end{align}
In two dimensions (2D), $\xi$ represents the angle between the $x$ axis and the normal, and uniformly covers the solid angle. In three dimensions (3D), $\zeta$ is the angle between the $z$ axis and the normal, and $\xi$ is the rotation of $\bn$ around the $z$ axis - see Fig.~\ref{fig:angles}b. For 3D isotropic systems, $\xi$ must also be uniform over the interval from 0 to $2\pi$ and $\zeta$ has the following probability density function
\begin{align}
f_{\xi}(\xi) = \begin{cases} \displaystyle\frac{1}{2\pi} & \mathrm{for}\ \xi\in[0, 2\pi] \\ 0 & \mathrm{otherwise}\end{cases}  \quad\quad
f_{\zeta}(\zeta) = \begin{cases} \displaystyle\frac{\sin\zeta}{2} & \mathrm{for}\ \zeta\in[0,\pi] \\ 0 & \mathrm{otherwise}\end{cases}
\end{align}

The second fundamental vector governing the behavior of the contact is the contact vector $\bt$. It is defined relative to the normal vector $\bn$ by angles $\chi$  and $\theta$ - see Fig.~\ref{fig:angles}b. The requirement of isotropicity dictates that (i) in 2D $\chi$ must have probability density function symmetric around zero and (ii) in 3D $\theta$ must be uniformly distributed over the 0--2$\pi$ interval.
\begin{align}
f_{\theta}(\theta) &= \begin{cases} \displaystyle\frac{1}{2\pi} & \mathrm{for}\ \theta\in[0, 2\pi] \\ 0 & \mathrm{otherwise}\end{cases} 
\end{align}
The probability distribution $f_{\chi}$ can be arbitrary (but symmetric around zero in 2D). For the sake of simplicity, it will now be assumed that the maximum angle between  $\bn$ and $\bt$ is $\gamma\in\left[ 0,\, \pi \right]$ and that all directions within this range are equally probable. 
\begin{align}
\mathrm{in\ 2D:\ } f_{\chi}(\chi) &= \begin{cases} \displaystyle\frac{1}{2\gamma} & \mathrm{for}\ \theta\in[-\gamma, \gamma] \\ 0 & \mathrm{otherwise}\end{cases} &
\mathrm{in\ 3D:\ }f_{\chi}(\chi) &= \begin{cases} \displaystyle\frac{\sin\chi}{1-\cos\gamma} & \mathrm{for}\ \theta\in[0, \gamma] \\ 0 & \mathrm{otherwise}\end{cases}  \label{eq:chi_restriction}
\end{align}
This assumption will later be removed in Sec.~\ref{sec:arbitrary_chi}.

Let us define a~second order tensor (rotation matrix) that provides the following relation between $\bn$ and $\bt$
\begin{align}
\bt = \bR\cdot\bn \label{eq:tRn} 
\end{align}
In two dimensions, this tensor is the standard rotation matrix
\begin{align}
\bR(\chi) = \left[\begin{array}{cc} \cos\chi & -\sin\chi\\ \sin\chi & \cos\chi \end{array}\right]
\end{align}
A~more complex situation is in 3D. One can imagine the construction of $\bn$ by taking the vector $\left(\begin{array}{ccc}0&0&1\end{array}\right)$, rotating it along the $y$ axis by angle $\zeta$ and then along the $z$ axis by angle $\xi$ (Fig.~\ref{fig:angles}b). In the same way, the construction of $\bt$ is achieved via four successive rotations along axes $y$, $z$, $y$ and $z$ by angles $\chi$, $\theta$, $\zeta$ and $\xi$, respectively.
\begin{align}
\bn &= \bR_z(\xi)\cdot\bR_y(\zeta)\cdot\left(\begin{array}{ccc}0&0&1\end{array}\right) & \bt &= \bR_z(\xi)\cdot\bR_y(\zeta)\cdot\bR_z(\theta)\cdot\bR_y(\chi)\cdot\left(\begin{array}{ccc}0&0&1\end{array}\right)
\end{align}
The rotation matrix from Eq.~\eqref{eq:tRn} is therefore
\begin{align}
\bR(\xi,\zeta,\chi,\theta) = \bR_z(\xi)\cdot\bR_y(\zeta)\cdot\bR_z(\theta)\cdot\bR_y(\chi)\cdot\bR_y^T(\zeta)\cdot\bR_z^T(\xi)
\end{align}

The cosine of angle $\chi$ between $\bn$ and $\bt$ can be calculated using Eq.~\eqref{eq:tRn}
\begin{align}
\cos\chi = \bn\cdot\bt = \bn\cdot\bR\cdot\bn = \bR:\left(\bn \otimes\bn \right) = \bR:\bN \label{eq:cospsi}
\end{align}
where the second order tensor $\bN$ is defined according to \citet{KuhDadd-00} as $\bN=\bn\otimes \bn$. Based on the assumption of no gaps or overlapping between the rigid bodies of the model, the volume of the domain is a~summation over the volumes of individual mechanical elements
\begin{align}
V = \displaystyle\sum\limits_e V_e = \displaystyle\sum\limits_e \cos\chi_e\dfrac{A_e l_e}{\Ndim} = \displaystyle\sum\limits_e \bR_e:\bN_e\dfrac{A_e l_e}{\Ndim}
\label{eq:volume}
\end{align}
where the number of dimensions $\Ndim$ is 2 or 3. Note that the volume of an~individual element is negative if $|\chi|>\pi/2$.

\section{Equivalence of virtual work}
The fundamental assumption about system degrees of freedom is taken from~\citep{KuhDadd-00}. It is known as Voigt's hypothesis and widely used as a~homogenization technique (e.g.~\cite{MikJir17}). It is assumed that when a~discrete system is subjected to constant strain $\bm{\varepsilon}$, all the rotations are zero and differences in translations are dictated by differences in position
\begin{align} 
\varphi&=0 &\bm{u}_b-\bm{u}_a &=  \bm{\varepsilon} \cdot \left(\bm{x}_b-\bm{x}_a\right) \label{eq:homdef}
\end{align}
The displacement jump at the contact between cells $a$ and $b$ is, based on the previous assumption, given by the rigid body kinematics of bodies without rotation
\begin{align}
\bm{\Delta} &= \bm{u}_b-\bm{u}_a =  l \bm{\varepsilon} \cdot \bt \label{eq:delta}
\end{align} 
where $l$ and $\bm{t}$ are respectively the length and contact vector belonging to the element connecting bodies $a$ and $b$. The normal and shear strain directly follow
\begin{align}
e_N &=\frac{\bn\cdot\bm{\Delta}}{l} = \bn \cdot \bm{\varepsilon} \cdot \bt  &  \bm{e}_T &=  \frac{\bm{\Delta} }{l}- e_N\bn = \bm{\varepsilon} \cdot \bt - \left(\bn \cdot \bm{\varepsilon} \cdot \bt\right) \bn \label{eq:strainTheoretical}
\end{align}
The stresses read
\begin{align}
s_N &= E_0 e_N  &  \bm{s}_T &= E_0\alpha\bm{e}_T \label{eq:facetStress}
\end{align}
where $E_0$ is the normal stiffness coefficient and $\alpha$ is the tangential/normal stiffness ratio. Both of the material parameters are considered constant throughout the whole domain.

The virtual work done by a~single element is obtained via integration of the product of the stress and the displacement jump (both constant) over the contact face 
\begin{align}
\delta W &= \int\limits_A (s_N\bn + \bm{s}_T)\cdot\delta\bm{\Delta} \dd{A} =  Al \left(s_N \delta e_N +  \bm{s}_T \cdot \delta \bm{e}_T\right) \label{eq:workSingleE1}
\end{align}
The integration is simple because the assumed zero rotations imply constant stresses and the displacement jump over the whole contact face. The total virtual work in the discrete system is the summation of these individual contributions.

To simplify the notation, we introduce the transpose operation $T_{ij}$ on an~arbitrary tensor $\bm{A}$ of sufficient order by swapping indices $i$ and $j$. 
\begin{align}
\bm{A}_{\dots i \dots j\dots}^{T_{ij}} = \bm{A}_{\dots j \dots i\dots}
\end{align}

Let us now define two additional tensors according to \citet{KuhDadd-00}: the fourth order tensor $\bIvol$ and the third order tensor $\bT$.
\begin{align}
\bIvol&=\frac{\bm{1}\otimes\bm{1}}{3}\\
\bT&= 3\bn\cdot \left(\bIvol\right)^{T_{13}} - \bn\otimes \bn \otimes \bn
\end{align}
where $\bm{1}$ is the identity matrix of size $N_{\mathrm{dim}}$. Note that $\bT$ is different from the definition in \cite{KuhDadd-00,Eli17} because the symmetry implied by equality $\bt=\bn$ is no longer present. The transposition $T_{13}$ means that dimensions 1 and 3 are swapped. 
 
With the previously defined tensors $\bN$ and $\bT$, Eq.~\eqref{eq:strainTheoretical} can be rewritten as
\begin{align}
e_N  &= \left(\bN \cdot \bR^T\right) : \bm{\varepsilon}  &  \bm{e}_T &=  \left(\bT \cdot\bR^T\right) : \bm{\varepsilon}  \label{eq:facetStrain}
\end{align}
using the transposition $T$ of the second order tensor swapping its two dimensions $T=T_{12}$.  The virtual work of a~single element (Eq.~\eqref{eq:workSingleE1}) can be rewritten as well.
\begin{align}
\delta W &= \nonumber A l \left(s_N \delta e_N +  \bm{s}_T \cdot \delta \bm{e}_T\right)\\
&=  A l E_0 \left(\left[\left(\bN\cdot \bR^T\right):\bm{\varepsilon}\right]\left[ \left(\bN \cdot \bR^T\right) : \delta\bm{\varepsilon} \right] +   \alpha\left[ \left(\bT\cdot \bR^T\right):\bm{\varepsilon}\right]\cdot\left[\left(\bT \cdot\bR^T\right) : \delta\bm{\varepsilon}\right]\right)
\nonumber\\&=  A l E_0 \left[\bm{\varepsilon}:\left(\bR\cdot\bN\otimes\bN\cdot\bR^T\right)^{T_{12}}:\delta\bm{\varepsilon} + \alpha \bm{\varepsilon}:\left(\bR\cdot\bT^{T_{13}}\cdot\bT\cdot\bR^T\right):\delta\bm{\varepsilon} \right] \nonumber\\ &= A l E_0 \bm{\varepsilon}: \left(\left(\bR\cdot\bN\otimes\bN\cdot\bR^T\right)^{T_{12}} + \alpha \bR\cdot\bT^{T_{13}}\cdot\bT\cdot\bR^T \right):\delta\bm{\varepsilon}\\
&= \nonumber A l E_0 \bm{\varepsilon}: \left(\pN + \alpha\pT  \right):\delta\bm{\varepsilon}
\end{align}
where
\begin{align}
\pN&=\left(\bR\cdot\bN\otimes\bN\cdot\bR^T\right)^{T_{12}} &
\pT&=\bR\cdot\bT^{T_{13}}\cdot\bT\cdot\bR^T
\end{align}

The total virtual work of the discrete assembly is 
\begin{align}
\delta W^{\mathrm{dis}} = \sum\limits_e \delta W_e  = \sum\limits_e A_e l_e E_0 \bm{\varepsilon}: \left(\pN_e + \alpha \pT_e \right):\delta\bm{\varepsilon}   \label{eq:virtW_discrete} 
\end{align}

The virtual work of an~equally strained elastic isotropic Boltzmann continuum occupying the same volume $V$ is 
\begin{align}
\delta W^{\mathrm{con}} = V\bm{\sigma}:\delta\bm{\varepsilon} = V\bm{\varepsilon}:\bD:\delta\bm{\varepsilon}
\label{eq:virtW_homog} 
\end{align}
with the constitutive equation $\bm{\sigma}=\bD:\bm{\varepsilon}$ where $\bD$ is fourth order tensor of elastic constants.

The equivalence of the discrete and continuous system implies the equality of virtual works (Hill-Mandel condition \cite{Hil63})
\begin{align}
\delta W^{\mathrm{dis}} = \delta W^{\mathrm{con}} \label{eq:VWequivalence}
\end{align}
Substituting Eqs.~\eqref{eq:virtW_discrete} and \eqref{eq:virtW_homog} into Eq.~\eqref{eq:VWequivalence}, the following expression for the tensor of elastic constants is derived
\begin{align}
\bD = \left\langle\frac{1}{V}\sum\limits_e A_e l_e E_0 \left(\pN_e + \alpha \pT_e  \right)\right\rangle^{\mathrm{SYM}} \label{eq:D_initial}
\end{align}
The symmetrization is needed because the tensors $\pN$ and $\pT$ do not possess the symmetries required for the Boltzmann continuum, which are \emph{major} symmetry (derived from the equivalence of mixed derivatives of elastic potential) and \emph{minor} symmetry (derived from symmetry of stress and strain tensors $\sigma_{ij} = \sigma_{ji}$,   $\varepsilon_{ij} = \varepsilon_{ji}$). 
\begin{align}
\mathrm{major\ symmetry:\quad}\cD_{ijkl} &= \cD_{klij} & \mathrm{minor\ symmetry:\quad}\cD_{ijkl} &=  \cD_{jikl} = \cD_{ijlk} = \cD_{jilk}
\end{align}
Because of the non-symmetric stress tensor in the discrete system, the minor symmetry is violated. The symmetric part can be easily obtained using transposition $T_{34}$.
\begin{align}
\left\langle\bullet\right\rangle^{\mathrm{SYM}} = \frac{\bullet+\bullet^{T_{34}}}{2} \label{eq:symmetrization}
\end{align}

Thanks to the assumed statistical independence between the normal and contact vector and the elemental area and length, the summation in Eq.~\eqref{eq:D_initial} can be broken into the following expression
\begin{align}
\bD = \frac{E_0}{V}\left\langle\mathrm{E}\left[\pN\right] + \alpha \mathrm{E}\left[\pT\right]\right\rangle^{\mathrm{SYM}}\sum\limits_e A_e l_e \label{eq:D}
\end{align}
where $\mathrm{E}\left[\bullet(\bm{x})\right]$ is the mean value of function $\bullet$, which is dependent on vector $\bm{x}$ with the distribution function $f_{\bm{X}}(\bm{x})$
\begin{align}
\mathrm{E}\left[\bullet(\bm{x})\right] = \int\limits_{-\infty}^{\infty} \dots \int\limits_{-\infty}^{\infty} \bullet(\bm{x}) f_{\bm{X}}(\bm{x}) \dd{\bm{x}}
\end{align}

Substituting $V$ from Eq.~\eqref{eq:volume} and utilizing the statistical independence again, one obtains
\begin{align}
\bD = \frac{\Ndim E_0}{\mathrm{E}[\bR:\bN]}\left\langle\mathrm{E}\left[\pN\right] + \alpha \mathrm{E}\left[\pT\right]\right\rangle^{\mathrm{SYM}} \label{eq:D0}
\end{align}

\section{Calculation of expectations}
\subsection{Two dimensional case}
In two dimensions, the rotation matrix depends only on angle $\chi$, while the normal $\bn$ depends only on angle $\xi$. The calculation of the mean value can be separated into two steps. Let us first calculate all the quantities dependent solely on $\bn$. 
\begin{align}
\mathrm{E}\left[\bN\right] &= \int\limits_0^{2\pi} \bn\otimes\bn \frac{1}{2\pi} \dd{\xi} = \frac{1}{2}\bm{1}
\\
\mathrm{E}\left[\bN\otimes\bN\right] &= \int\limits_{0}^{2\pi} \bN\otimes\bN \frac{1}{2\pi}\dd{\xi}= \frac{1}{4}\bI +  \frac{3}{8}\bIvol 
\\
\mathrm{E}\left[\bT^{T_{13}}\cdot\bT\right] &= \int\limits_{0}^{2\pi} \bT^{T_{13}}\cdot\bT \frac{1}{2\pi}\dd{\xi}= \frac{3}{4}\bI -  \frac{3}{8}\bIvol - \frac{3}{2}\left(\bIvol\right)^{T_{23}} 
\end{align}
where the fourth order tensor  $\bI=\mathscr{I}_{ijkl}=(\delta_{ik}\delta_{jl}+\delta_{il}\delta_{jk})/2$ delta is employed;  $\delta_{ij}\equiv\bm{1}$ is the the Kronecker delta. 

In the second step, theses quantities are used in the calculation of the mean values of terms involving both $\bR$ and $\bn$.
\begin{align}
\mathrm{E}\left[\bR:\bN\right] &= \frac{1}{2}\int\limits_{-\gamma}^{\gamma} \bR:\bm{1} \frac{1}{2\gamma}\dd{\chi} = \frac{1}{4\gamma}\int\limits_{-\gamma}^{\gamma} 2\cos\chi \dd{\chi} = \frac{\sin\gamma}{\gamma} \label{eq:vol2D}
\\
\mathrm{E}\left[\pN\right] =&  \nonumber \int\limits_{-\gamma}^{\gamma}\int\limits_{0}^{2\pi} \pN  \frac{1}{2\gamma} \frac{1}{2\pi} \dd{\xi}\dd{\chi} = \int\limits_{-\gamma}^{\gamma} \left[\bR\cdot\mathrm{E}\left[\bN\otimes\bN\right]\cdot\bR^T\right]^{T_{12}} \frac{1}{2\gamma} \dd{\chi}\\
=& \int\limits_{-\gamma}^{\gamma}\left[\bR\cdot\left(\frac{1}{4}\bI +  \frac{3}{8}\bIvol\right)\cdot\bR^T\right]^{T_{12}} \frac{1}{2\gamma} \dd{\chi} \\
=& \nonumber \frac{3}{4} \left(\bIvol\right)^{T_{23}} + \frac{3\sin2\gamma}{16\gamma}\left( 
\bIvol - \left(\bIvol\right)^{T_{23}} + \left(\bIvol\right)^{T_{24}}\right)
\\
\mathrm{E}\left[\pT\right] =&  \nonumber \int\limits_{-\gamma}^{\gamma}\int\limits_{0}^{2\pi}  \pT \frac{1}{2\gamma} \frac{1}{2\pi} \dd{\xi}\dd{\chi} = \int\limits_{-\gamma}^{\gamma} \bR\cdot\mathrm{E}\left[\bT^{T_{13}}\cdot\bT\right]\cdot\bR^T \frac{1}{2\gamma} \dd{\chi}\\
=& \int\limits_{-\gamma}^{\gamma}\bR\cdot\left(\frac{3}{4}\bI -  \frac{3}{8}\bIvol - \frac{3}{2}\left(\bIvol\right)^{T_{23}}\right)\cdot\bR^T \frac{1}{2\gamma} \dd{\chi} \\
=& \nonumber \frac{3}{4} \left(\bIvol\right)^{T_{24}} - \frac{3\sin2\gamma}{16\gamma}\left( 
\bIvol + \left(\bIvol\right)^{T_{23}} - \left(\bIvol\right)^{T_{24}}\right)
\end{align}

Only the symmetric parts of these expectations are needed. The following identities, which are valid for both the 2D and the 3D model, are derived from Eq.~\eqref{eq:symmetrization} 
\begin{align}
\left\langle\left(\bIvol\right)^{T_{23}}\right\rangle^{\mathrm{SYM}} = \left\langle\left(\bIvol\right)^{T_{24}}\right\rangle^{\mathrm{SYM}} = \frac{\bI}{3} \label{eq:symI}
\end{align}
and used to obtain the final symmetric expectations of tensors.
\begin{align}
\left\langle\mathrm{E}\left[\pN\right]\right\rangle^{\mathrm{SYM}} &=\frac{\mathrm{E}\left[\pN\right] + \mathrm{E}\left[\pN\right]^{T_{34}}}{2} = \frac{1}{4}\bI + \frac{3\sin2\gamma}{16\gamma} \bIvol  \label{eq:bigN2D} \\ 
\left\langle\mathrm{E}\left[\pT\right]\right\rangle^{\mathrm{SYM}} &=\frac{\mathrm{E}\left[\pT\right] + \mathrm{E}\left[\pT\right]^{T_{34}}}{2} = \frac{1}{4}\bI - \frac{3\sin2\gamma}{16\gamma} \bIvol \label{eq:bigT2D}
\end{align}

\subsection{Three dimensional case}

In three dimensions, integration is substantially more complex. It is performed over four independent angles and cannot be separated since rotation matrix $\bR$ depends on all four angles. Calculation by hand is extremely tedious; it was performed by computer instead. The following three integrations were delivered with the~help of the Python library for symbolic mathematics, SymPy \cite{SymPy}.  
\begin{align}
\mathrm{E}\left[\bR:\bN\right] &= \int\limits_{-\gamma}^{\gamma}\int\limits_{0}^{2\pi}\int\limits_{0}^{\pi}\int\limits_{0}^{2\pi} \bR:\bN  \frac{1}{2\pi} \frac{\sin\zeta}{2} \frac{1}{2\pi} \frac{\sin\chi}{1-\cos\gamma}  \dd{\xi}\dd{\zeta}\dd{\theta}\dd{\chi} = \cos^2\left(\frac{g}{2}\right) \label{eq:vol3D}
\\
\mathrm{E}\left[\pN\right] =&  \nonumber \int\limits_{-\gamma}^{\gamma}\int\limits_{0}^{2\pi}\int\limits_{0}^{\pi}\int\limits_{0}^{2\pi} \pN  \frac{1}{2\pi} \frac{\sin\zeta}{2} \frac{1}{2\pi} \frac{\sin\chi}{1-\cos\gamma}  \dd{\xi}\dd{\zeta}\dd{\theta}\dd{\chi} = \\
=& \frac{1}{3} \left(\bIvol\right)^{T_{23}} + \frac{2\cos\gamma + \cos 2\gamma + 1}{20}\left( 
\bIvol + \left(\bIvol\right)^{T_{24}} - \frac{2}{3} \left(\bIvol\right)^{T_{23}}\right)
\\
\mathrm{E}\left[\pT\right] =&  \nonumber \int\limits_{-\gamma}^{\gamma}\int\limits_{0}^{2\pi}\int\limits_{0}^{\pi}\int\limits_{0}^{2\pi} \pT  \frac{1}{2\pi} \frac{\sin\zeta}{2} \frac{1}{2\pi} \frac{\sin\chi}{1-\cos\gamma}  \dd{\xi}\dd{\zeta}\dd{\theta}\dd{\chi} = \\
=& \frac{2}{3} \left(\bIvol\right)^{T_{24}} - \frac{2\cos\gamma + \cos 2\gamma + 1}{20}\left( 
\bIvol + \left(\bIvol\right)^{T_{23}} - \frac{2}{3} \left(\bIvol\right)^{T_{24}}\right)
\end{align}

Using identity \eqref{eq:symI}, the symmetric part yields
\begin{align}
\left\langle\mathrm{E}\left[\pN\right]\right\rangle^{\mathrm{SYM}} &=\frac{\mathrm{E}\left[\pN\right] + \mathrm{E}\left[\pN\right]^{T_{34}}}{2} = 
\frac{2\cos\gamma + \cos 2\gamma + 21}{180} \bI + \frac{2\cos\gamma + \cos 2\gamma + 1}{20} \bIvol \label{eq:bigN3D}
\\
\left\langle\mathrm{E}\left[\pT\right]\right\rangle^{\mathrm{SYM}} &=\frac{\mathrm{E}\left[\pT\right] + \mathrm{E}\left[\pT\right]^{T_{34}}}{2} = \frac{39 - 2\cos\gamma - \cos 2\gamma}{180} \bI - \frac{2\cos\gamma + \cos 2\gamma + 1}{20} \bIvol \label{eq:bigT3D}
\end{align}

\section{Relation between the elastic parameters of discrete system and continuum}

The mechanical behavior of a~linearly elastic isotropic solid is determined by two constants. Here we choose the elastic modulus ($E$) and Poisson's ratio ($\nu$). The tensor of the elastic constants is expressed using these variables
\begin{align}
\bD = \begin{cases}\displaystyle\frac{E}{1+\nu}\bI+\frac{3E\nu}{1-\nu^2}\bIvol & \mathrm{2D,\,plane\ stress} \\[7pt]
\displaystyle\frac{E}{1+\nu}\bI+\frac{3E\nu}{(1+\nu)(1-2\nu)}\bIvol & \mathrm{2D,\, plane\ strain} \\[7pt]
\displaystyle\frac{E}{1+\nu}\bI+\frac{3E\nu}{(1+\nu)(1-2\nu)}\bIvol &\mathrm{3D}
\end{cases} \label{eq:Dmacro}
\end{align}
Equation~\eqref{eq:D0}, along with symmetrized expectations  \eqref{eq:vol2D}, \eqref{eq:bigN2D} and \eqref{eq:bigT2D} in two dimensions as well as \eqref{eq:vol3D}, \eqref{eq:bigN3D} and \eqref{eq:bigT3D} in three dimensions, provides 
\begin{align}
\bD = \begin{cases}
\displaystyle E_0\left[(1+\alpha)\frac{\gamma}{2\sin\gamma}\bI+(1-\alpha)\frac{3\cos\gamma}{4}\bIvol\right] & \mathrm{2D} \\[7pt]
\displaystyle{E_0}\left[
\dfrac{(1-\alpha)(2\cos\gamma+\cos(2\gamma)-39)+60}{30(\cos\gamma+1)}
\bI + \frac{3(1-\alpha)}{5}\cos\gamma \bIvol\right] &\mathrm{3D}
\end{cases} \label{eq:Dmeso}
\end{align}

\begin{figure}[tb!]
\centering\includegraphics[width=\textwidth]{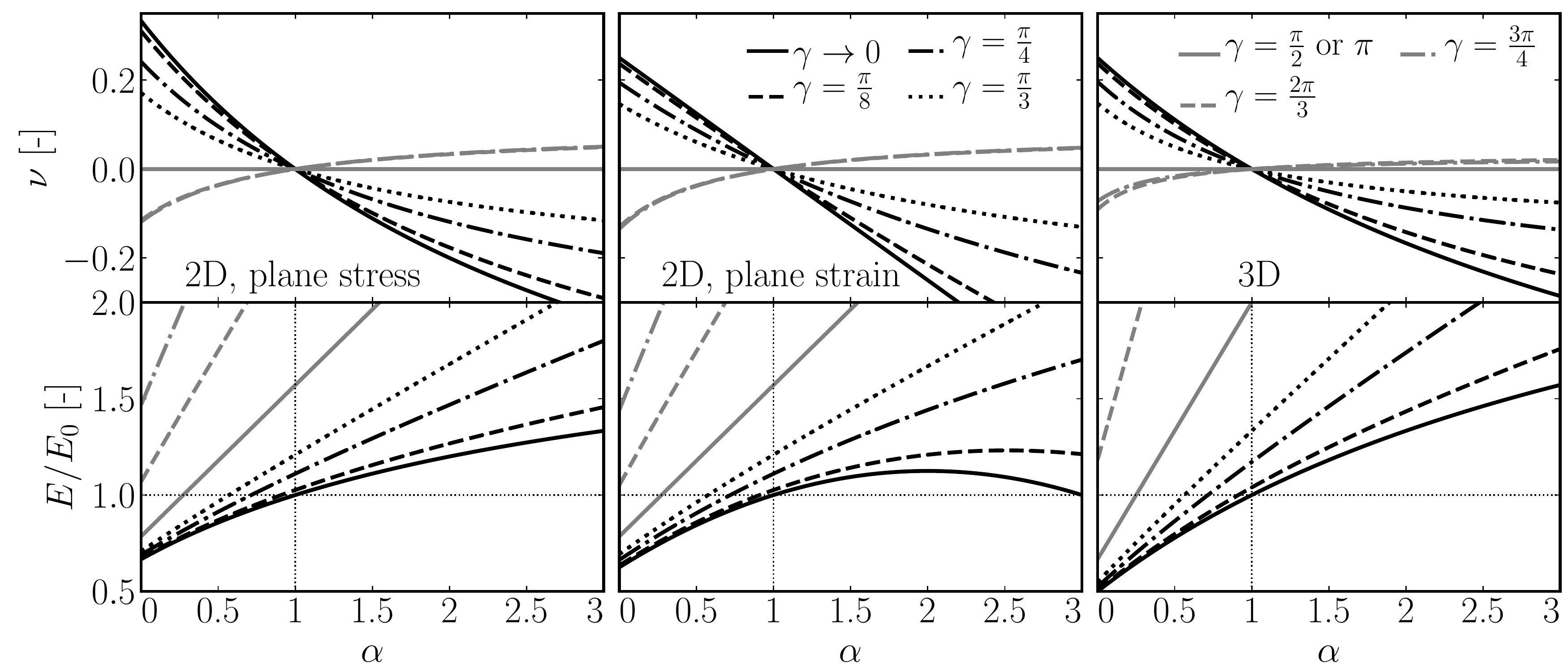}
\caption{Macroscopic elastic properties of two and three dimensional discrete systems with all directions of $\bt$ equally probable (limited by $|\chi|<\gamma$) dependent on the tangential/normal stiffness ratio $\alpha$ according to Eqs.~\eqref{eq:nu_gamma} and \eqref{eq:E_gamma}.} 
\label{fig:elastic_constants_gamma}
\end{figure}

Equality between meso and macroscopic elastic tensors \eqref{eq:Dmeso} and \eqref{eq:Dmacro} requires equality between the respective scalar multipliers of tensors $\bIvol$ and $\bI$. One can modify these algebraic equations into relations between macroscopic parameters $E$ and $\nu$ and mesoscopic parameters $E_0$, $\alpha$ and $\gamma$.  
\begin{align}
\nu&=\begin{cases}\dfrac{\left(1-\alpha\right)\sin2\gamma}{4(1+\alpha)\gamma +(1-\alpha)\sin2\gamma} & \mathrm{2D,\,plane\ stress}\\[8pt]
\dfrac{\left(1-\alpha\right)\sin2\gamma}{4(1+\alpha)\gamma +2(1-\alpha)\sin2\gamma} &\mathrm{2D,\,plane\ strain} \\[8pt]
\dfrac{3(1-\alpha)(\cos\gamma+\cos^2(\gamma))}{(1-\alpha)(7\cos\gamma+7\cos^2\gamma-20)+30} & \mathrm{3D}
\end{cases}
\label{eq:nu_gamma}\\
E&=\begin{cases}E_0\dfrac{2(1+\alpha)^2 \gamma^2 + (1-\alpha^2)\gamma\sin 2\gamma}{\sin\gamma(4(1+\alpha)\gamma + (1-\alpha)\sin 2\gamma)} & \mathrm{2D,\,plane\ stress}\\[8pt]
E_0\dfrac{4(1+\alpha)^2 \gamma^2 + 3(1-\alpha^2)\gamma\sin 2\gamma}{\sin\gamma(8(1+\alpha)\gamma + 4(1-\alpha)\sin 2\gamma} &\mathrm{2D,\,plane\ strain} \\[8pt]
E_0\dfrac{2\left[(1-\alpha)(\cos\gamma+\cos^2\gamma-20)+30\right]\left[(1-\alpha)(\cos\gamma+\cos^2\gamma-2)+3\right]}{(1-\alpha)(7\cos\gamma+7\cos^2\gamma-20)+30} & \mathrm{3D}
\end{cases}
\label{eq:E_gamma}
\end{align}
These equations are plotted in Fig.~\ref{fig:elastic_constants_gamma} for the range $\alpha\in[0,\,3]$.

Decreasing $\gamma$ towards zero must yield relations for a~discrete system with $\bn=\bt$. 
\begin{align}
\lim\limits_{\gamma\rightarrow 0}\nu&=\begin{cases}  \dfrac{1-\alpha}{3+\alpha} & \mathrm{2D,\,plane\ stress} \\[7pt] \dfrac{1-\alpha}{4} & \mathrm{2D,\,plane\ strain} \\[7pt] \dfrac{1-\alpha}{4+\alpha} & \mathrm{3D} \end{cases}
&
\lim\limits_{\gamma\rightarrow 0}E&=\begin{cases}  E_0\dfrac{2+2\alpha}{3+\alpha} & \mathrm{2D,\,plane\ stress} \\[7pt]  E_0\dfrac{(1+\alpha)(5-\alpha)}{8} & \mathrm{2D,\,plane\ strain} \\[7pt] E_0\dfrac{2+3\alpha}{4+\alpha} & \mathrm{3D} \end{cases} \label{eq:lim_gamma_0}
\end{align}
Indeed, the calculation of limits provides correct expressions (derived for example in \citep{BatRot88,LiaCha-97,Eli17} under the assumption of the perpendicularity of the contact vector and contact face). They are also identical to those from microplane theory \cite{CarBaz97}.

\begin{figure}[tb!]
\centering\includegraphics[width=\textwidth]{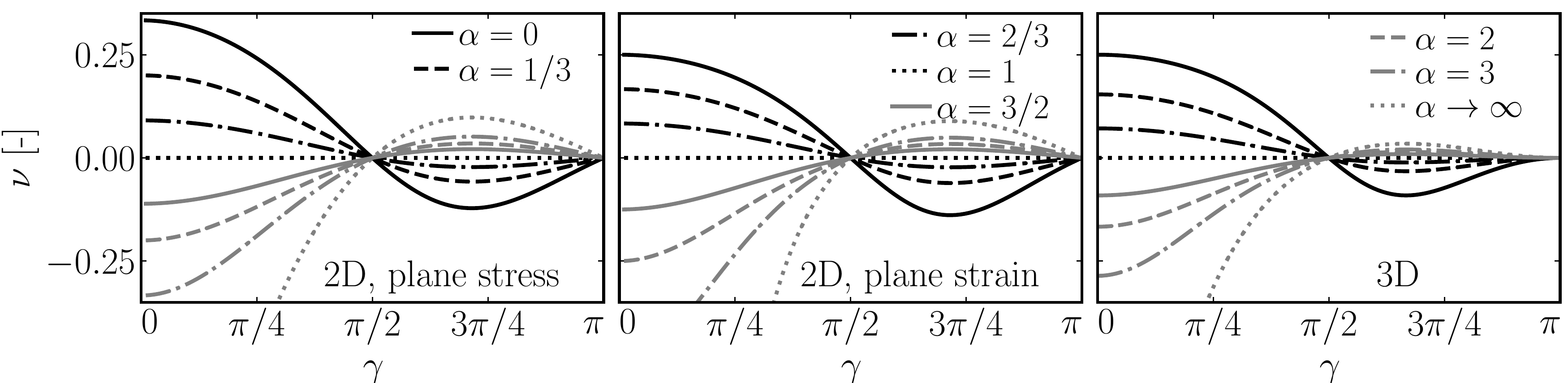}
\caption{The macroscopic Poisson's ratio of two and three dimensional discrete systems with all direction of $\bt$ equally probable (limited by $|\chi|<\gamma$) dependent on the limit $\gamma$ according to Eqs.~\eqref{eq:nu_gamma} and \eqref{eq:E_gamma}.} 
\label{fig:elastic_constants_gamma_gamma}
\end{figure}

What are the maximum and minimum Poisson's ratios that can be achieved? One can differentiate the expression with respect to $\gamma$ and search for a~stationary point (leaving out the degenerative case $\alpha=1$). In 2D, such an~analysis reveals a~local extreme at points $\gamma=0$ and $\gamma\approx2.24670$ (solution of $2\gamma=\tan2\gamma$). In 3D, the stationary points are $\gamma=0$, $\pi$ and  $\approx2.09440$ (exactly $\arccos (-0.5)$). Plotting the Poisson's ratio with respect to the limit angle $\gamma$ (Fig.~\ref{fig:elastic_constants_gamma_gamma}) shows that the maximum range of $\nu$ is obtained for $\gamma=0$, i.e. when the contact vector equals the normal vector. This is the classic solution stated in Eq.~\eqref{eq:lim_gamma_0}. Increasing $\gamma$ up to $\pi/2$ causes the interval of achievable Poisson's ratios to shrink to zero. Then, the interval opens again with opposite signs; its width maximizes at $\gamma=2.24670$ (2D) or $\gamma=2.09440$ (3D). The interval of possible values of $\nu$ for these $\gamma$ reads:  $\left[-0.122,\,0.098\right]$ for 2D plane stress, $\left[-0.139,\,0.089\right]$ for 2D plane strain and $\left[-0.091,\,0.034\right]$ for 3D, respectively.
These maximum and minimum value of Poisson's ratio occur always for $\alpha=0$ or $\alpha\rightarrow\infty$.

We have proven that under assumption~\eqref{eq:chi_restriction}, one cannot increase the Poisson's ratio limits beyond that which is provided by the model with $\bn=\bt$ in Eq.~\eqref{eq:lim_gamma_0}. 

\section{Arbitrary distribution $f_{\chi}(\chi)$ \label{sec:arbitrary_chi}}
 This section proves that the same conclusion unfortunately holds for the arbitrary distribution of angle $\chi$. The only restriction on the probability density of $\chi$ applied here comes from the isotropy requirement in 2D which demands that $f_{\chi}(\chi)$ be symmetric around zero. In 3D, $f_{\chi}(\chi)$ can be arbitrary.

\begin{figure}[tb!]
\centering\includegraphics[width=\textwidth]{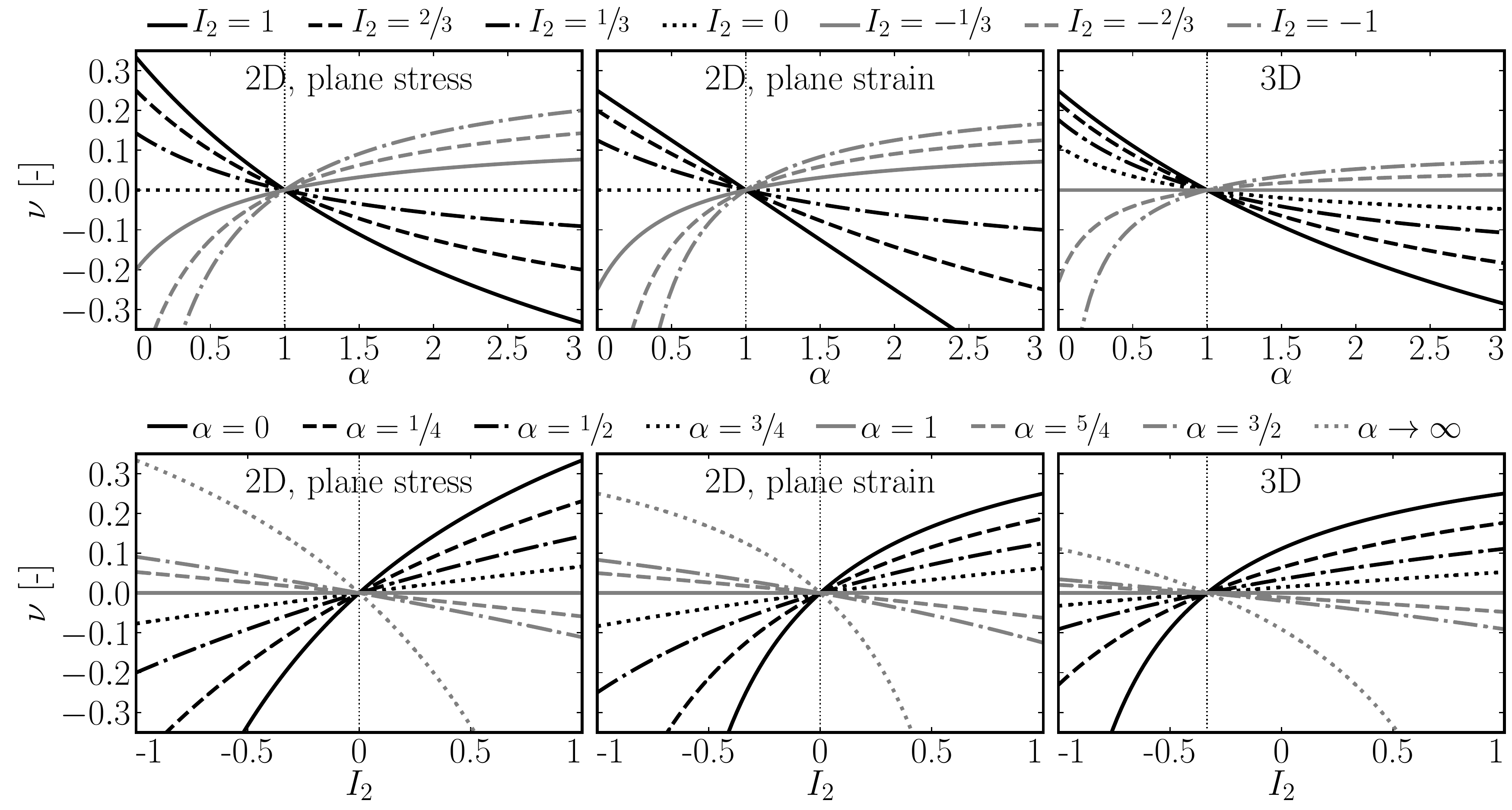}
\caption{Poisson's ratio of two and three dimensional discrete systems with arbitrary distribution of angle $\chi$ between the normal and contact vector according to Eq.~\eqref{eq:arbit_nu}. The variable $I_2$ represents integral $\int_{\Omega_{\chi}} \cos 2\chi f_{\chi}(\chi)\dd{\chi}$.} 
\label{fig:elastic_constants_I}
\end{figure}

Let us denote the following auxiliary integrals
\begin{align}
\int\limits_{\Omega_{\chi}} \cos \chi \,f_{\chi}(\chi) \dd{\chi} &= I_1 & \int\limits_{\Omega_{\chi}} \cos(2\chi) f_{\chi}(\chi) \dd{\chi} &= I_2 
\label{eq:aux_integrals}
\end{align}
$\Omega_{\chi}$ is the domain of the probability distribution function $f_{\chi}(\chi)$ which is the interval $[-\pi,\,\pi ]$ in 2D and $[ 0,\,\pi ]$ in 3D. With the help of the Python symbolic mathematics library SymPy \cite{SymPy}, the expectations needed in Eq.~\eqref{eq:D0} were computed. Using identity  $\int_{\Omega_{\chi}} f_{\chi}(\chi) \dd{\chi}=1$ resulting from the definition of probability density; and in 2D $\int_{\Omega_{\chi}} \sin(2\chi) f_{\chi}(\chi) \dd{\chi}=0$ derived from the symmetry of $f_{\chi}$, the expectations are
\begin{align}
\left\langle\mathrm{E}\left[\pN\right] + \alpha\mathrm{E}\left[\pT\right]\right\rangle^{\mathrm{SYM}}  & = \begin{cases}\dfrac{1+\alpha}{4}\bI + \dfrac{3(1-\alpha)I_2}{8}\bIvol & \mathrm{2D}\\
\dfrac{(1-\alpha)(I_2-13)+20}{60}\bI + (1-\alpha)\dfrac{3I_2+1}{20}\bIvol & \mathrm{3D}
\end{cases} \label{eq:arbit_expectations1}\\
\mathrm{E}\left[\bR:\bN\right] &= I_1 \label{eq:arbit_expectations2}
\end{align}
Substituting these expressions into Eq.~\eqref{eq:D0}, requiring equality with Eq.~\eqref{eq:Dmacro} and solving for unknown $E$ and $\nu$, provides 
\begin{align}
\nu &= \begin{cases}\dfrac{(1-\alpha)I_2}{2(1+\alpha)+(1-\alpha)I_2} & \mathrm{2D,\,plane\  stress}\\[8pt]
\dfrac{(1-\alpha)I_2}{2(1+\alpha)+2(1-\alpha)I_2} & \mathrm{2D,\,plane\  strain}\\[8pt]
\dfrac{(1-\alpha)(3I_2+1)}{(1-\alpha)(7I_2-11)+20} & \mathrm{3D}\end{cases} 
\label{eq:arbit_nu}
\\
E &= \begin{cases}E_0\dfrac{(1+\alpha)^2 + (1-\alpha^2)I_2}{(2(1+\alpha)+(1-\alpha)I_2)I_1}& \mathrm{2D,\,plane\  stress}\\[8pt]
E_0\dfrac{2(1+\alpha)^2 + 3(1-\alpha^2)I_2}{4(1+\alpha+(1-\alpha)I_2)I_1}&  \mathrm{2D,\,plane\  strain}
\\[8pt]E_0\dfrac{((1-\alpha)(I_2-13)+20)(I_2(1-\alpha) + 1+\alpha)}{2((1-\alpha)(7I_2-11)+20)I_1}&  \mathrm{3D}\end{cases}
\label{eq:arbit_E}
\end{align}
The Poisson's ratio predicted by Eq.~\eqref{eq:arbit_nu} is plotted in Fig.~\ref{fig:elastic_constants_I}. 

Let us focus again on the theoretical limits of Poisson's ratio provided by Eq.~\eqref{eq:arbit_nu}. The meaningful values of $\alpha$ range from 0 to $\infty$. The integral $I_2$ is limited to an~interval from -1 to 1 because these are the maximum and minimum values of the continuous function $\cos(2\chi)$, which in the integral is ``weighted'' by an~arbitrary non-negative function with the unit integral over its domain. 
The only way $I_2=1$ can be obtained is when $f_{\chi}(\chi)$ is a~discrete distribution with zero probability everywhere except points $\chi=\pi$, 0 or $-\pi$ (the last of which is possible only in 2D). All of these angles imply that $\bt$ is parallel to $\bn$. The second limit case of $I_2=-1$ is only possible when $f_{\chi}(\chi)$ is zero everywhere except points $\chi=\pm\pi/2$, i.e. when $\bt$ is perpendicular to $\bn$.

Differentiating Eq.~\eqref{eq:arbit_nu} with respect to $I_2$ reveals that, except in the degenerative case of $\alpha=1$, there is no stationary point within the investigated domain for both 2D and 3D models. The extreme values therefore lie on the boundaries. One can see in Figure~\ref{fig:elastic_constants_I} that the maximum value of $\nu$ is reached by minimizing $\alpha$ and maximizing $I_2$ or by maximizing $\alpha$ and minimizing $I_2$; the opposite is true for the opposite goal of reaching the minimum Poisson's ratio.

In 2D, the same Poisson's ratio limits are obtained for both $I_2=1$ or $I_2=-1$, and these limits are the same as those of the model where $\bn=\bt$. Indeed, the equality $\bn=\bt$ implies $I_2=1$, and substituting that into Eq.~\eqref{eq:arbit_nu} yields  Eq.~\eqref{eq:lim_gamma_0}. In 3D, the case $I_2=1$ obtained for parallel $\bn$ and $\bt$ again yields Eq.~\eqref{eq:lim_gamma_0} and the widest range of Poisson's ratio. The case with perpendicular $\bn$ and $\bt$ (when $I_2=-1$), leads in 3D to a~narrower interval as well as any other distribution of $\chi$.  

We have proven that an~isotropic discrete structure with an~arbitrary relation between $\bn$ and $\bt$ cannot increase the Poisson's ratio limits beyond the limits obtained by the model with $\bn=\bt$ in Eq.~\eqref{eq:lim_gamma_0}. 

\section{Macroscopic elastic properties of actual discrete systems}

Let us now observe the behavior of actual discrete systems and compare it to our analytical formulas. We divide a~domain into (generally non-convex) polygons or polyhedrons; this tessellation defines the shapes of rigid bodies. Contact forces $\bF=A(s_N\bn+\bm{s}_T)$ arise at the centroids of the edges (or faces) of polygons (or polyhedrons, respectively). We search for displacements and rotations fulfilling linear and angular momentum balance equations. These equations, for a~single discrete body without external load, read
\begin{align}
\sum \limits_e \bF_e = E_0\sum \limits_e A_e \left(e_N^e \bm{n}_e + \alpha \bm{e}_T^e \right) &= \bm{0} \label{eq:linear_momentum}\\
\sum \limits_e  \bF_e\cdot\bE\cdot\bm{r}_e = E_0 \sum \limits_e A_e \left(e_N^e \bm{n}_e + \alpha \bm{e}_T^e \right)\cdot\bE\cdot\bm{r}_e &= \bm{0} \mathrm{\ in\ 3D\ or\ } 0 \mathrm{\ in\ 2D} \label{eq:angular_momentum}
\end{align}
where $e$ runs over all contacts with body neighbors, $\bm{r}$ is a~position vector of the contact force $\bF$ and $\bE$ is the Levi-Civita permutation symbol. In 3D, the contraction of $\bE$ from both sides gives the vector product $\bm{b}\cdot\bE\cdot\bm{a} = \bm{a}\times\bm{b}$.

We limit this section to 2D models, as it is expected that 3D models would yield similar results. Four different 2D tessellation types are considered, namely the \emph{Voronoi}, \emph{randomized Voronoi}, \emph{random} and \emph{centered random} tessellation.    

\subsection{Voronoi tessellation}
The first tessellation type, referred to as \emph{Voronoi} hereinafter, has parallel normal and contact vectors. It can be obtained via Voronoi tessellation, which is widely used in discrete modeling, or via Power tessellation, which is capable of taking into account the size of the inclusions (mineral aggregates in concrete) associated with rigid bodies \citep{Eli16}.

The \emph{Voronoi} model is created here by placing points randomly into a~domain in sequence and accepting only those with a~minimum distance from previously placed points that is greater than the length parameter $\lmin$. The sequential placement process is terminated after no point is accepted for a~sufficiently large number of trials. The random points serve as nuclei for clipped Voronoi tessellation and their translations and rotations constitute model degrees of freedom. 

The macroscopic elastic behavior of the \emph{Voronoi} model is described by Eq.~\eqref{eq:lim_gamma_0}. Numerical verification is performed in \citep{Eli16} for both the 2D and 3D models, revealing increasing deviation from Eq.~\eqref{eq:lim_gamma_0} with the increasing distance of parameter $\alpha$ from 1. The deviation is caused by the violation of assumption~\eqref{eq:homdef}, which is exactly met only for $\alpha=1$. Indeed, assuming $\alpha=1$, applying Eqs.~\eqref{eq:delta} and \eqref{eq:strainTheoretical} that were derived from assumption~\eqref{eq:homdef}, and using $\bt=\bn$, which is valid for \emph{Voronoi} tessellation, the equilibrium equations become
\begin{align}
E_0\sum \limits_e A_e \frac{\bm{\Delta}_e}{l_e} = E_0\bm{\varepsilon}\cdot\sum \limits_e A_e \bm{n}_e =E\bm{\varepsilon}\cdot\int \limits_\Gamma \bm{n} \dd{\Gamma} &= 0 \label{eq:lin_mom_Voronoi}\\
E_0\sum \limits_e A_e \frac{\bm{\Delta}_e}{l_e}\cdot\bE\cdot\bm{r}_e= E_0\bE:\left[\left(\sum \limits_e A_e \bm{r}_e \otimes \bm{n}_e\right) \cdot \bm{\varepsilon}\right] = E_0\bE:\left[\int \limits_\Gamma \bm{r} \otimes \bm{n} \dd{\Gamma} \cdot \bm{\varepsilon}\right] &=  \bm{0} \mathrm{\ or\ } 0  \label{eq:ang_mom_Voronoi}
\end{align}
The sum over the contacts is transformed into integration over the enclosed surface $\Gamma$ of the rigid body. The first integral is the zero vector, while the second integral is the identity matrix multiplied by the rigid body volume. This is derived via component-wise integration with the help of the divergence theorem and unit standard Cartesian basis vector $\bm{j}$.
\begin{align}
\int \limits_\Gamma n_j \dd{\Gamma} = \int \limits_\Gamma \bm{j}\cdot \bn \dd{\Gamma} = \int \limits_V \nabla\cdot\bm{j}\dd{V} &= 0 \\
\int \limits_\Gamma r_i n_j \dd{\Gamma} = \int \limits_\Gamma r_i\,\bm{j}\cdot\bn \dd{\Gamma} = \int \limits_V \nabla\cdot(r_i\,\bm{j})\dd{V} = \int \limits_V \frac{\partial r_i}{\partial x_j}\dd{V}  &= \begin{cases} V & \mathrm{for\ } i=j \\ 0
& \mathrm{for\ } i\neq j\end{cases}  \label{eq:angular_momentum_mod} \end{align}
Since the first integration is the zero vector, the right-hand side of Eq.~\eqref{eq:lin_mom_Voronoi} is zero and Eq.~\eqref{eq:linear_momentum} is exactly satisfied. Substituting the result of the second integration, the right-hand side of Eq.~\eqref{eq:angular_momentum_mod} becomes $VE_0\bE:\bm{\varepsilon}$, which is always zero thanks to the symmetry of the strain tensor and the antisymmetry of the Levi-Civita symbol, and Eq.~\eqref{eq:angular_momentum} is exactly satisfied as well.

We have shown that for \emph{Voronoi} tessellation with $\alpha=1$, assumption~\eqref{eq:homdef} holds. However, for different $\alpha$ or non-parallel $\bn$ and $\bt$, assumption~\eqref{eq:homdef} is incorrect. The actual system is more compliant and has a~higher Poisson's ratio than that predicted by assumption~\eqref{eq:homdef} due to the removed constraint on rotations and translations.

\begin{figure}[tb!]
\centering\includegraphics[width=13cm]{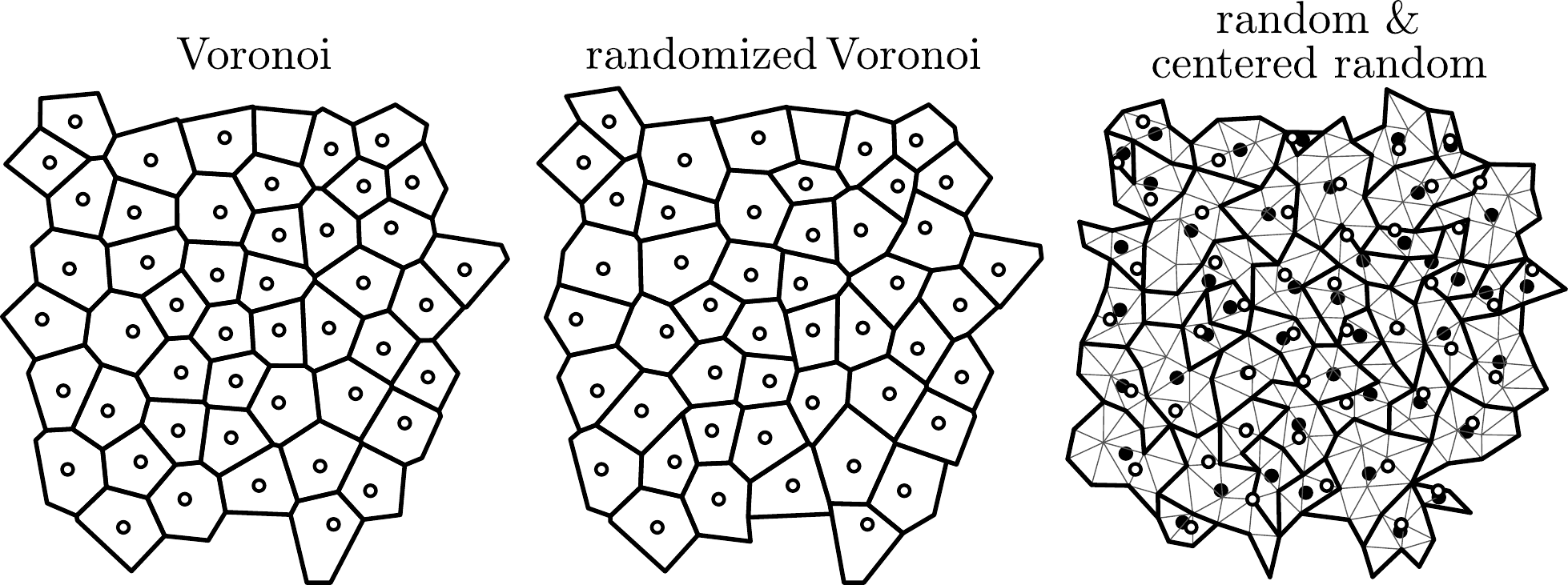}
\caption{Four types of tessellation created with the same set of nuclei (hollow circles). The centroids of the bodies of \emph{random tessellation} are plotted by solid circles.} 
\label{fig:tessellations}
\end{figure}

\begin{figure}[tb!]
\centering\includegraphics[width=8cm]{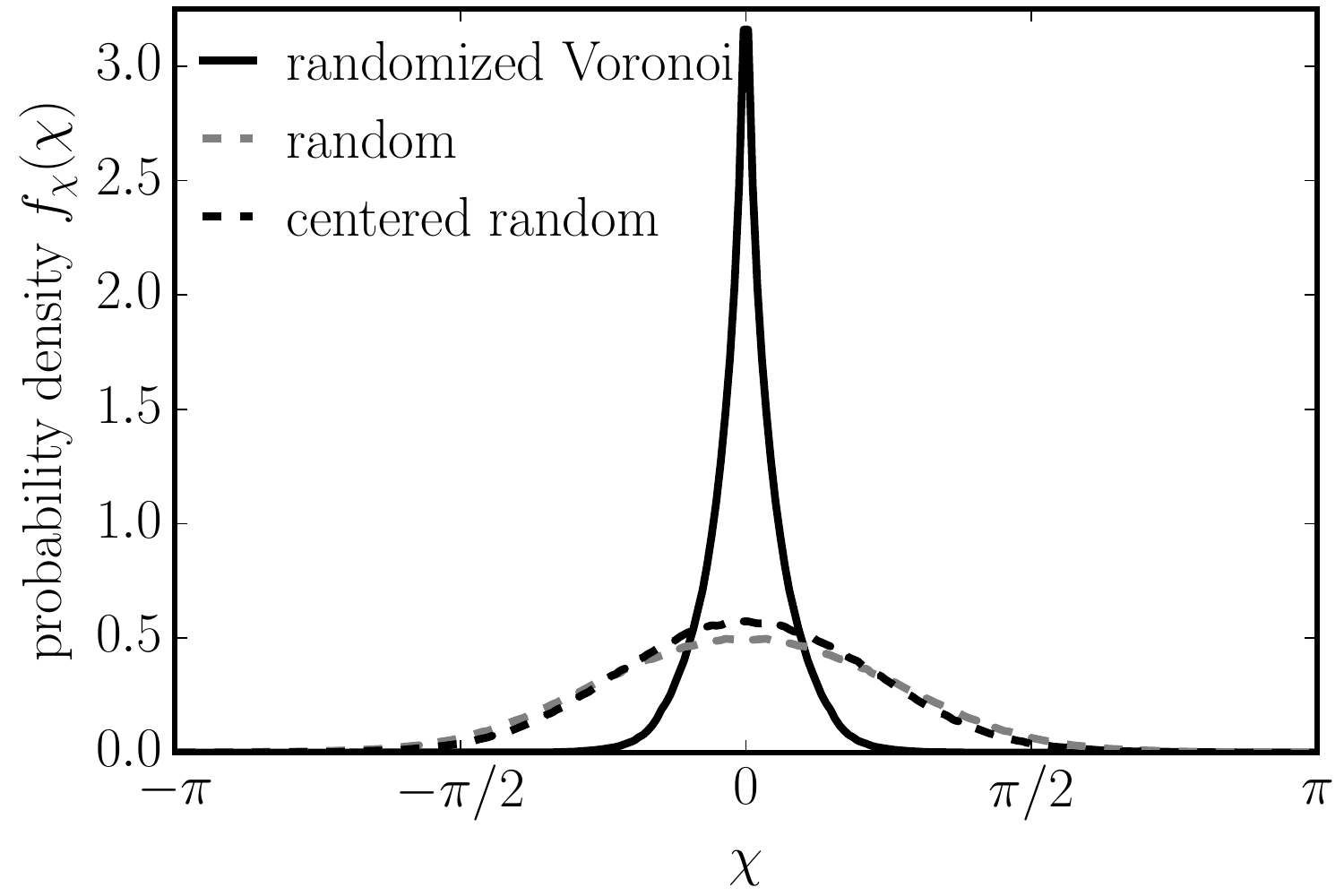}
\caption{Probability density function $f_{\chi}(\chi)$ for \emph{randomized Voronoi}, \emph{random} and \emph{centered random} tessellation. 
\label{fig:chist_chi}}
\end{figure}

\subsection{Randomized Voronoi tessellation \label{sec:rand_vor}}

\emph{Randomized Voronoi} tessellation is generated by modifying the \emph{Voronoi} model. We simply move each Voronoi vertex in a~random direction by a~random distance. The normal vectors, $\bm{n}$, are randomly rotated by such random movements, while the contact vectors, $\bm{t}$, remain intact.  The random distance is generated from the uniform distribution in the interval $(0,\,k)$, where $k$ is half of the minimum distance to the closest vertex. This upper limit on $k$ is introduced to prevent overlapping of the rigid bodies. The resulting tessellation has generally non-convex body shapes and non-parallel $\bm{n}$ and $\bm{t}$. 

By generating 50 discrete structures of size $150\lmin\times150\lmin$, the distribution function of angle $\chi$ is obtained. Numerical integration of Eq.~\eqref{eq:aux_integrals} then provides constants $I_1\approx0.97724$ and $I_2\approx0.91372$. An~example of \emph{randomized Voronoi} tessellation  as well as the probability distribution function of $\chi$ is shown in Figs.~\ref{fig:tessellations} and \ref{fig:chist_chi}.

\subsection{Random tessellation}

The randomization of the Voronoi structure defined in section~\ref{sec:rand_vor} is still rather limited and strongly resembles the original \emph{Voronoi} model. When searching for a~tessellation with more deviation from the parallelism of $\bt$ and $\bn$, the following process was found to be quite effective. The \emph{random} model is based on two sets of nodes: basic nodes, which are generated by the same sequential process as described above with the minimum distance $\lmin$, and auxiliary nodes (vertices), which are created independently in the same way with the minimum distance $\lmin/2$. Delaunay triangulation is performed on the vertices, and triangles containing a~basic node are directly assigned to it. Then, an~iterative loop is performed over all unassigned triangles. Whenever a~neighboring triangle already belonging to a~basic node is found, the unassigned triangle is assigned to the same basic node. The rigid body associated with a~given basic node is then a~union of all the triangles assigned to that node. The obtained shapes are highly non-convex and typically several contact faces are created between two neighboring bodies.

Again, the distribution function of angle $\chi$ is estimated from 50 discrete structures of the size $150\lmin\times150\lmin$. This is shown along with an~example of \emph{random} tessellation in Figs.~\ref{fig:tessellations} and \ref{fig:chist_chi}. Numerical integration of Eq.~\eqref{eq:aux_integrals} provides $I_1\approx0.73516$ and $I_2\approx0.28884$.

\subsection{Centered random tessellation}

The last tessellation type has bodies which are identical to those in the \emph{random} model. The difference is that for \emph{centered random} tessellation, a~final step is performed in which the governing nodes bearing the rigid body degrees of freedom are moved into the centroids of the generated bodies. The situation is depicted in Fig.~\ref{fig:tessellations}, where the hollow circles (initial tessellation nuclei) are replaced by solid circles (centroids). Though the shape of the bodies is unchanged, the geometrical characteristics of the tessellation changes because the vectors $\bt$ and lengths $l$ are updated. The statistical evaluation of angle $\chi$ on 50 discrete structures with a~size of $150\lmin\times150\lmin$ provides the probability distribution function shown in Fig. \ref{fig:chist_chi} that, when numerically integrated in Eq.~\eqref{eq:aux_integrals}, provides $I_1\approx0.78830$ and $I_2\approx0.38688$.

One would assume that since the rigid body shapes are equal for the \emph{random} and \emph{centered random} model, these models would behave equally. We will now prove the opposite by contradiction.   

Let us assume two structures with equal shapes and connectivity of the rigid bodies but two different sets of governing nodes (with coordinates $\bm{x}_i$ and $\hat{\bm{x}}_i$, respectively)  which bear their degrees of freedom. The normals $\bn$ and contact areas $A$ remain the same because these are dictated solely by rigid body shape. However, the contact lengths and contact vectors are different, $l \neq \hat{l}$ and $\bt\neq\hat{\bt}$. The starting point of the proof by contradiction is that under equal load the movements of the bodies are equal as well, and therefore the displacement jumps at the boundaries are identical, $\bm{\Delta} = \hat{\bm{\Delta}}$. According to the first parts of Eqs.~\eqref{eq:strainTheoretical} combined with Eq.~\eqref{eq:facetStress}, the stresses differ by factor $l/\hat{l}$ 
\begin{align}
\hat{s}_N &= E_0 \frac{\bn \cdot \bm{\Delta}}{\hat{l}} = s_N\frac{l}{\hat{l}} &
\hat{\bm{s}}_T &= E_0\alpha\left(\frac{\bm{\Delta}}{\hat{l}} - \hat{e}_N\bn\right) = E_0\alpha\left(\frac{\bm{\Delta}}{\hat{l}} - \frac{\bn \cdot \bm{\Delta}}{\hat{l}}\bn\right) = \bm{s}_T\frac{l}{\hat{l}}
\end{align}
The same factor also holds for contact forces as these are only stresses multiplied by areas: $\hat{\bF} = l/\hat{l}\bF $

The first set of forces ($\bF$) satisfies equilibrium equations \eqref{eq:linear_momentum} and  \eqref{eq:angular_momentum} because we assume it originates from the true solution with the first set of nuclei. We will show now that the second set ($\hat{\bF}$) violates equilibrium equations because the factor $l/\hat{l}$ differs for each element
\begin{align}
\sum \limits_e \hat{\bF}_e =\sum \limits_e \bF_e \frac{l_e}{\hat{l}_e} &\neq \sum \limits_e \bF_e  = \bm{0}\\
\sum \limits_e  \hat{\bF}_e\cdot\bE\cdot\bm{r}_e = \sum \limits_e  \frac{l_e}{\hat{l}_e} \bF_e\cdot\bE\cdot\bm{r}_e &\neq \sum \limits_e  \bF_e\cdot\bE\cdot\bm{r}_e = \bm{0} \mathrm{\ or\ } 0
\end{align}
The assumption of the equality of rigid body movements for different governing nodes leads to a~contradiction. Therefore, the movements are generally different and the macroscopic elastic properties of the \emph{random} and \emph{centered random} model differ as well. 

\subsection{Comparison of analytical formulas to actual characteristics}
A~two dimensional model with a~size of $150\lmin\times150\lmin$ was generated for all four tessellation types. The structure was loaded by the prescribed translations and rotations along the whole boundary, $\Gamma$: $u^{\Gamma}_1=p x_1$, $u^{\Gamma}_2=q x_2$ and $\varphi^{\Gamma}=0$. The resulting macroscopic strain components are $\varepsilon_{11}=p$, $\varepsilon_{22}=q$ and $\varepsilon_{12}=\varepsilon_{21}=0$, respectively. Alternatively, the strain tensor can be obtained via the linear regression of location-translation dependence, see \citep{Eli16}. The stress tensor was estimated from all inner contacts with a~distance from the boundary greater than $3\lmin$ \citep{Bag96}
\begin{align}
\bm{\sigma} = \left\langle\sum\limits_e \bF_e \otimes \bm{c}_e\right\rangle^{\mathrm{SYM}}
\end{align}
where $\bm{c}$ is the centroid of the contact face and the symmetrization of the second order tensor reads $\langle \bullet \rangle^{\mathrm{SYM}} = (\bullet + \bullet^{T})/2$. 

Since the diagonal strain term  $\varepsilon_{12}$ is zero, the  macroscopic elastic properties are easily evaluated 
\begin{align}
\nu &= \begin{cases} \dfrac{\sigma_{22}\varepsilon_{11}-\sigma_{11}\varepsilon_{22}}{\sigma_{11}\varepsilon_{11}-\sigma_{22}\varepsilon_{22}} & \mathrm{plane\ stress}\\[3mm]
\dfrac{\sigma_{22}\varepsilon_{11}-\sigma_{11}\varepsilon_{22}}{(\sigma_{11}+\sigma_{22})(\varepsilon_{11}-\varepsilon_{22})} & \mathrm{plane\ strain}
\end{cases} \label{eq:poisson_numeric}
\\
E &= \begin{cases}\dfrac{\sigma_{11}^2-\sigma_{22}^2}{\sigma_{11}\varepsilon_{11}-\sigma_{22}\varepsilon_{22}} & \mathrm{plane\ stress}\\[3mm] \dfrac{(\sigma_{11}-\sigma_{22})(\varepsilon_{11}(\sigma_{11}+2\sigma_{22})-\varepsilon_{22}(2\sigma_{11}+\sigma_{22}))}{(\varepsilon_{11}-\varepsilon_{22})^2(\sigma_{11}+\sigma_{22})} & \mathrm{plane\ strain}
\end{cases} \label{eq:E_numeric}
\end{align}

\begin{figure}[tb!]
\centering\includegraphics[width=15cm]{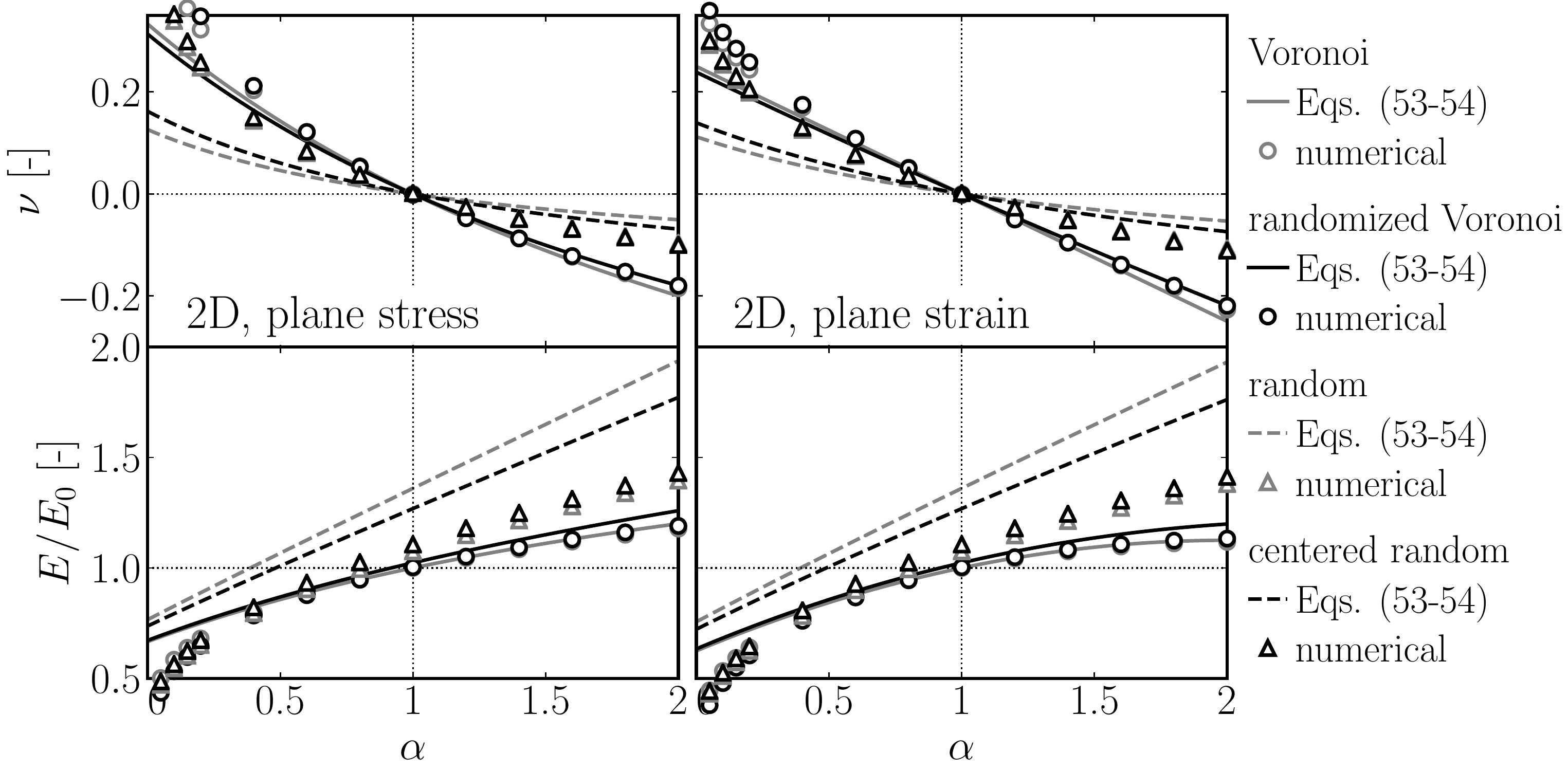}
\caption{The macroscopic elastic characteristics (Poisson's ratio and elastic modulus) of a~two dimensional discrete system derived analytically and computed numerically for four different tessellation types.} 
\label{fig:elastic_constants_NUM}
\end{figure}

The model was assembled and solved for each tessellation type and various $\alpha$ ratios. The macroscopic characteristics obtained by Eqs.~\eqref{eq:poisson_numeric} and \eqref{eq:E_numeric} are plotted in Fig.~\ref{fig:elastic_constants_NUM} along with the derived analytical predictions \eqref{eq:arbit_nu} and \eqref{eq:arbit_E}. There is reasonable correspondence for \emph{Voronoi} and \emph{random Voronoi} tessellation when $\alpha>0.3$. In all the other cases the predictions severely departed from numerical solution due to the unfulfilled fundamental assumption \eqref{eq:homdef}. However, the elastic tensor evaluated using Eq.~\eqref{eq:D_initial} on the generated structure correspond in all the cases with one from Eq.~\eqref{eq:D0} when analytical expectations \eqref{eq:arbit_expectations2} and \eqref{eq:arbit_expectations2} are used.

Though the correspondence is rather weak, there is a~clear trend seen in both the analytical and the numerical results. The Poisson's ratio shrinks into a~narrower interval as the integral $I_2$ grows due to the loss of parallelism between $\bn$ and $\bt$; simultaneously, the elastic modulus increases. We consider this result to be a~verification of our conclusion that Poisson's ratio limits are maximized for Voronoi (or Power) tessellation. Any departure from perpendicularity between $\bt$ and the face plane causes the narrowing of these limits. 

\section{Conclusions}

Analytical estimations are derived for the macroscopic elastic behavior of (i) discrete isotropic assemblies with (ii) normal and shear force linearly dependent on normal and shear displacement discontinuity and (iii) a~continuously filled domain. The derivation takes advantage of a~strong assumption about the rotations (assumed zero everywhere) and translations (assumed uniformly distributed over domain) of discrete bodies. Comparison with numerical results reveals that the assumption is only reasonable for a~limited range of model types, though the overall trend emerging from the derived equations is confirmed.  
\begin{itemize}
\item It is proven that the widest limits of Poisson's ratio are obtained when contact and normal vectors are parallel, such as when Voronoi or Power tessellation is used to generate rigid body shapes. 
\item Any other model geometry (where normal and contact vectors are not strictly parallel) shrinks the interval of achievable Poisson's ratio. It is not possible to extend the limits via geometrical manipulations. These limits are $\nu\in\left[-1,\,\sfrac{1}{3}\right]$ for 2D plane stress, $\nu\in\left(-\infty,\,\sfrac{1}{4}\right]$ for 2D plane strain and $\nu\in\left[-1,\,\sfrac{1}{4}\right]$ for 3D models, respectively.
\item The mechanical parameters $E_0$ and $\alpha$ are considered constant throughout the whole volume. Also, no overlapping or gaps between rigid bodies are allowed. The abandonment of these conditions seems to be one of the possible ways to proceed further in searching for other methods of extending the Poisson's ratio interval of discrete models.
\item It is shown that the position of the governing node affects model results, even though the actual rigid bodies are identical. It is questionable whether the body centroids, Voronoi sites (generators of the Voronoi diagram) or other points should be used as the governing nodes. It would be advantageous to find a~model modification that would remove this dependence.
\end{itemize}

The discrete models are often used for analyzing inelastic phenomena (mostly fracture). 
The connection between the angle formed by the normal and contact vectors and the local and global inelastic behavior of the model is an interesting open topic deserving further investigation.

\section*{Acknowledgement}
Financial support provided by the Czech Science Foundation under project No. GA19-12197S is gratefully acknowledged.

\bibliographystyle{plainnat}
\bibliography{bibliography.bib}

\end{document}